%% file: 0-main.tex
\documentclass[sigconf,natbib=true]{acmart}

\AtBeginDocument{%
  \providecommand\BibTeX{{%
    \normalfont B\kern-0.5em{\scshape i\kern-0.25em b}\kern-0.8em\TeX}}}
    
\input{zhwinfonts}
\copyrightyear{2022} 
\acmYear{2022} 
\setcopyright{acmcopyright}\acmConference[CIKM '22]{Proceedings of the 31st ACM International Conference on Information and Knowledge Management}{October 17--21, 2022}{Atlanta, GA, USA}
\acmBooktitle{Proceedings of the 31st ACM International Conference on Information and Knowledge Management (CIKM '22), October 17--21, 2022, Atlanta, GA, USA}
\acmPrice{15.00}
\acmDOI{10.1145/3511808.3557670}
\acmISBN{978-1-4503-9236-5/22/10}

\usepackage[utf8]{inputenc}
\usepackage{graphicx}
\usepackage{amsmath}
\usepackage{algorithm}
\usepackage{algorithmic}
\usepackage{CJKutf8}
\usepackage{subcaption}
\usepackage{balance}
\usepackage{multirow}
\usepackage[skip=2.5pt]{caption}
\usepackage{array}
\usepackage{booktabs}
\usepackage{threeparttable}
\usepackage{soul}

\usepackage[compact]{titlesec}

\newcommand{\eg}{\textit{e.g.}}
\newcommand{\ie}{\textit{i.e.}}
\newcommand{\etc}{\textit{etc.}}
\newcommand{\R}{\mathbb{R}}
\newcommand\blfootnote[1]{%
  \begingroup
  \renewcommand\thefootnote{}\footnote{#1}%
  \addtocounter{footnote}{-1}%
  \endgroup
}
\begin{document}
% \fancyhead{}

\setlength{\abovedisplayskip}{2.5pt}
\setlength{\belowdisplayskip}{2.5pt}

\title{Pre-training Tasks for User Intent Detection and Embedding Retrieval in E-commerce Search}

\author{Yiming Qiu$^\dag$, Chenyu Zhao$^\dag$, Han Zhang, Jingwei Zhuo, Tianhao Li, Xiaowei Zhang, Songlin Wang, Sulong Xu, Bo Long and Wen-Yun Yang$\,^{*}$}
\affiliation{%
  \institution{JD.com, Beijing, China \& Shenzhen, China \& Mountain View, USA}
  \{\small qiuyiming3, zhaochenyu8, zhanghan33, zhuojingwei1, litianhao5, zhangxiaowei9, wangsonglin3, xusulong, bo.long, wenyun.yang\}@jd.com
  \country{}
}
\renewcommand{\shortauthors}{Y. Qiu, C. Zhao et al.}



\begin{abstract}
% Pretrain-Finetune paradigm recently becomes prevalent in many NLP tasks: question answering, text classification, sequence labeling and so on. As the state-of-the-art model, BERT pre-trained on the general corpus ({\eg}, Wikipedia) have been widely used in these tasks. 
% However, these BERT-style models still show limitations on some scenarios, especially for two: a corpus that contains very different text from the general corpus Wikipedia, or a task that learns embedding spacial distribution for specific purpose ({\eg}, approximate nearest neighbor search).
% In this paper, to tackle the above dilemmas we also encounter in an industrial e-commerce search system, we propose novel customized pre-training tasks for two critical modules: user intent detection and semantic embedding retrieval. 
% The customized pre-trained models with specific fine-tuning, being less than 10\% of BERT-base's size in order to be feasible for cost-efficient CPU serving, significantly improves its other counterparts on both offline evaluation metrics and online benefits.  We have open sourced our datasets~\footnote{https://github.com/jdcomsearch/jdsearch-22} for the sake of reproducibility and future works.
BERT-style models pre-trained on the general corpus (\eg, Wikipedia) and fine-tuned on specific task corpus, have recently emerged as breakthrough techniques in many NLP tasks: question answering, text classification, sequence labeling and so on. However, this technique may not always work, especially for two scenarios: a corpus that contains very different text from the general corpus Wikipedia, or a task that learns embedding spacial distribution for a specific purpose (\eg, approximate nearest neighbor search).
In this paper, to tackle the above two scenarios that we have encountered in an industrial e-commerce search system, we propose customized and novel pre-training tasks for two critical modules: user intent detection and semantic embedding retrieval.
The customized pre-trained models after fine-tuning, being less than 10\% of BERT-base's size in order to be feasible for cost-efficient CPU serving, significantly improve the other baseline models: 1) no pre-training model and 2) fine-tuned model from the official pre-trained BERT using general corpus, on both offline datasets and online system. We have open sourced our datasets~\footnote{https://github.com/jdcomsearch/jd-pretrain-data} for the sake of reproducibility and future works.

\blfootnote{$^\dagger\,$ Both authors contribute equally}
\blfootnote{$^*\,$ corresponding author}
\end{abstract}


\maketitle

\input{1-introduction}
\input{3-method}
\input{5-experiment}

\input{6-conclusion}

\bibliographystyle{ACM-Reference-Format}
\balance

\bibliography{references}

\end{document}

%% file: 1-introduction.tex
\section{Introduction}
\label{sec:introduction}

Over the recent decades, online shopping platforms (e.g., eBay, Walmart, Amazon, Tmall, Taobao and JD) have become increasingly popular in people's daily life. E-commerce search, which helps users find what they need from billions of products, is an essential part of those platforms, contributing to the largest percentage of transactions among all channels~\cite{Sorokina:2016:ASJ:2911451.2926725}. 
Nowadays, along with the recent advance in large-scale Nature Language Processing (NLP) pre-trained models, NLP plays an increasingly vital role in almost every module of e-commerce search. Thus, it is essential to develop powerful pre-trained NLP models to improve the overall performance of an e-commerce search system.

\begin{figure}
    \vspace{3mm}
    \centering
    \includegraphics [width=0.4\textwidth]{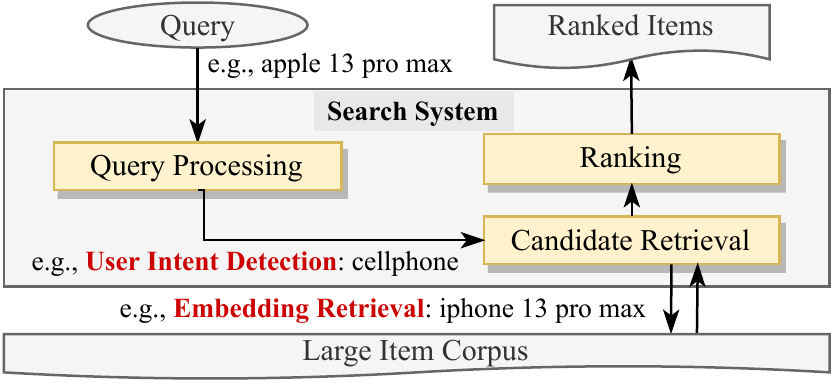}
    \caption{Major stages of an e-commerce search system.}
    \label{fig:search-engine}
    \vspace{-4mm}
\end{figure}

Figure~\ref{fig:search-engine} illustrates a typical workflow of an e-commerce search system, which includes query processing (including user intent detection), semantic retrieval, and ranking. Let's take the query ``apple 13 pro max'' as an example.
% 1) query auto-completion suggests ``apple 13 pro max'' when user types ``apple 13''. 
1) \underline{Query processing}, which aims to detect the query intentions like category, brand, {\etc}, recognizes the intention as ``cellphone''. With the great advances in NLP models, BERT-style models are massively used~\cite{cai2021slim, skinner2019commerce, lin2018commerce} in the task. 2) \underline{Candidate retrieval} is normally accomplished by traditional inverted index retrieval in the manner of keyword matching, while model-based semantic retrieval~\cite{huang2013learning, zhang2020towards, li2021embedding, liu2021que2search} methods are optimized for additional results which are semantically relevant with ``iphone 13 pro max''. 3) \underline{Ranking} model finally orders the retrieved candidates based on thousands of factors, such as relevance, user preference, product popularity, {\etc}

% With the recent significant advances in NLP models, various NLP techniques have been applied into search and information retrieval systems to achieve state-of-the-art performance~\cite{zhang2020towards,zhang2021joint,chang2019pre}. 
% Especially, BERT~\cite{devlin2018bert}, pre-trained on large-scale corpus such as Wikipedia and simply fine-tuned on domain-specific dataset, has already been proven to be the state-of-the-art model in many NLP tasks, such as question answering~\cite{yang2019end,wang2019multi}, text classification~\cite{sun2019fine,lu2020vgcn}, sequence labeling~\cite{tsai2019small,liu2019gcdt}, etc.
% % However, the above technique, though general enough, is not guaranteed to perform well for very specific domains. For example, researchers find the original MLM (masked language model) and NSP (next sentence prediction) pre-train tasks are not sufficient to train a best performing embedding retrieval model. % TODO: any other papers to discuss?
% Some works have been done to facilitate e-commerce search with BERT models. For example, \textit{Prod2BERT}~\cite{bianchi2021bert} learns product representation through masked session modeling based on BERT architecture. Another work uses a fine-tuned BERT model~\cite{jiang2020fine} to improve the quality of non-default ranking on e-commerce platform. While how to apply pre-trained models to other modules of e-commerce search remains a big challenge.

Recent significant advances in NLP modeling techniques provide us with a few promising directions~\cite{zhang2020towards,zhang2021joint,chang2019pre,chowdhery2022palm,raffel2019exploring}. 
Those large-scale models, such as BERT~\cite{devlin2018bert}, T5~\cite{raffel2019exploring}, PaLM~\cite{chowdhery2022palm} and so on, pre-trained on large-scale corpus such as Wikipedia and simply fine-tuned on domain-specific dataset, have already been proven to be the state-of-the-art model in many NLP tasks, such as question answering~\cite{yang2019end,wang2019multi}, text classification~\cite{sun2019fine,lu2020vgcn}, sequence labeling~\cite{tsai2019small,liu2019gcdt}, \etc, which inspire us to apply BERT-based pre-training models to the e-commerce search system.
In the view of NLP, nowadays e-commerce search system faces the following challenges that we are going to tackle in this paper.
\begin{itemize}
\item \underline{Free-form text} refers to the user typed texts (\eg, customer typed queries and merchant typed product descriptions) that does not follow grammar rules. 
On the one hand, the customer typed queries may be ambiguous for the system to identify exact user intent, since the system could be confused by word order (\eg, milk chocolate or chocolate milk), typos or unclear expressions. 
On the other hand, product descriptions typed by the merchant may be composed of a handful of short text segments that do not follow grammar strictly, like the product titles shown in Table~\ref{tab:title}.
As a result, BERT models pre-trained on Wikipedia or other corpus that follow grammar strictly directly on this problem are not suitable for this case. 
% We provide detailed analysis in Section \ref{sec:experiment}, in which we show applying pre-trained BERT models directly leads to deteriorated performance.
In Section~\ref{sec:experiment}, we will show the unsatisfied performance of the direct application of the pre-trained BERT models.
%Consequently, it is hard for us to directly apply the pre-trained BERT models, since it is usually trained by Wikipedia and other standard corpus that follow grammar strictly.
 
\item \underline{Long-tail query} has introduced additional difficulties in our model learning: 1) there are a lot of long-tail queries, and 2) the training examples for each long-tail query are far from enough. As a typical example, the query "Goliath", is a cold-start brand in our e-commerce platform. It is quite difficult to identify its product category intent and retrieve relevant products due to the lack of supervised training data. 
% Thus, recently this long-tail problem induces many seminal works:
Thus, several long-tail related studies have emerged: an across-context attention mechanism for long-tail query classification~\cite{zhang2021modeling}, a dual heterogeneous graph attention network to improve long-tail performance in e-commerce search~\cite{niu2020dual} and a systematical study of the long-tail effect in recommendation system~\cite{oestreicher2012recommendation}.

\end{itemize}
% In this paper, we take the pre-training approach to a more general solution to these problems, which facilitate two important components of a leading e-commerce search system, user intent detection in query processing step and semantic retrieval in candidate retrieval step, as shown in Figure~\ref{fig:search-engine}.
In this paper, we take the pre-training technology to realize a more general solution to these problems, especially for two important components of a leading e-commerce search system, {\ie} user intent detection in query processing step and semantic retrieval in candidate retrieval step, as shown in Figure~\ref{fig:search-engine}. We do not consider ranking module in this paper since it always tends to consider user’s personalized information. But we claim the proposed method is general enough to be applied in other modules in e-commerce search.

\begin{CJK*}{UTF8}{gbsn}
\linespread{1.2}
\begin{table}[t]
\centering
\caption{Typical e-commerce text examples that are repetitive, redundant and grammatically disastrous.}
\label{tab:title}
\footnotesize
\resizebox{\columnwidth}{!}{
\begin{tabular}{c}
\hline
\specialrule{0em}{2pt}{0pt}
{\normalsize 带芽人参种子人参苗子 西洋参种子 长白山人参种子药材种子}\\
(Ginseng seeds with buds Ginseng seedlings Western ginseng seeds Changbai Mountain ginseng seeds Medicinal seeds)
\\ \specialrule{0em}{0pt}{6pt}
{\normalsize鸡毛掸子家用车用鸡毛扫灰清洁用品汽车拖把不掉毛掸子除尘}\\
(Feather Duster Household Car Use Chicken Feather Dust Sweeping Cleaning Supplies Car Mop No Lint Duster Dust Removal)
\\ \specialrule{0em}{0pt}{6pt}
{\normalsize巴沙鱼片5斤装 冷冻鱼片 淡水龙利鱼柳 新鲜无刺无骨鱼肉做酸菜鱼 2500g} \\
(Basa Fish Fillet 5kg Frozen Fish Fillet Freshwater Longli Fish Fillet Fresh Boneless Fish Meat Sauerkraut Fish 2500g)
\\ \specialrule{0em}{2pt}{2pt}
\hline
\end{tabular}}
\vspace{-4mm}
\end{table}
\end{CJK*}

%% file: 3-method.tex
\section{Method}
\label{sec:method}

\subsection{User Intent Detection}
\label{sec:method:intent}
\subsubsection{Problem Formulation.}
\begin{figure}[t]
    \centering
    \includegraphics[width=0.48\textwidth]{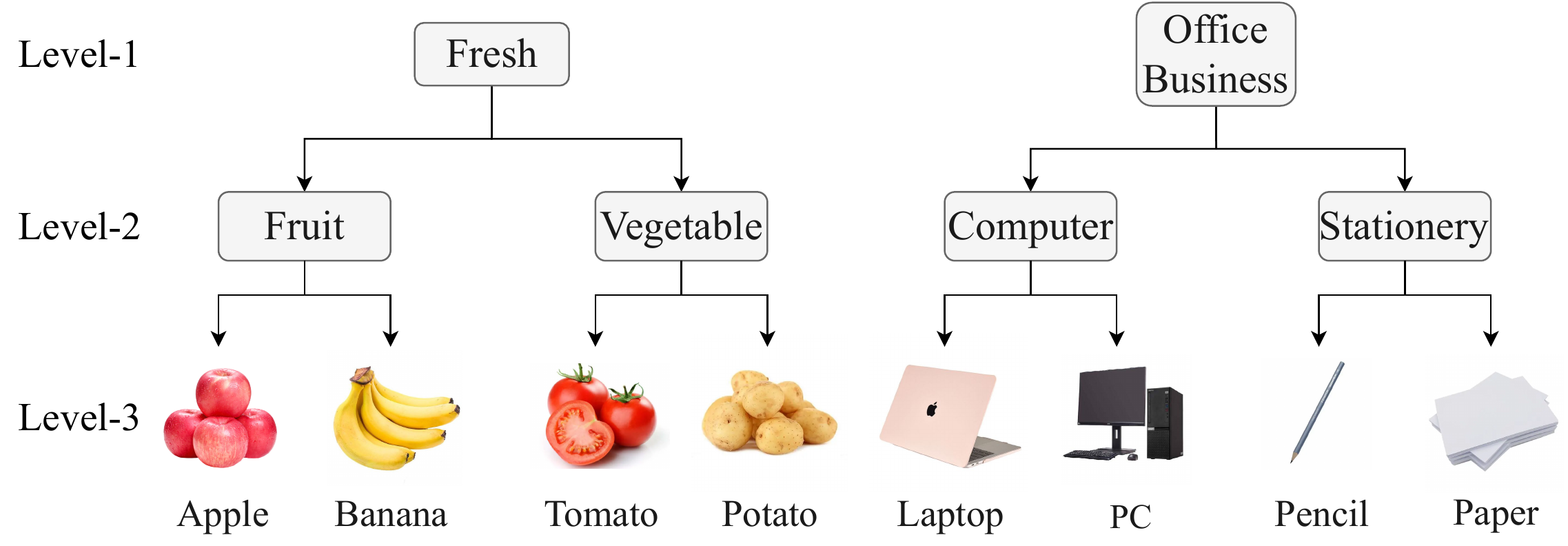}
    \caption{Product Category Hierarchy.}
    \label{fig:cate_hierarchy}
    \vspace{-5mm}
\end{figure}

User intent detection can be formulated as a standard multi-label classification problem where the total number of labels $L$ is usually large, \ie, 3,000, which equals the number of leaves in the hierarchical product category structure as shown in Figure~\ref{fig:cate_hierarchy}.
For a query $x$, our model learns a vector-valued output function $f(x) \in \mathbb{R}^{L}$ that produces the probability of each label. Then we obtain labels with top-$k$ probabilities, or set a probability threshold to get a dynamic number of predicted labels.

\subsubsection{Pre-training Tasks.}
\begin{figure*}[t]
    \centering
    \begin{subfigure}{0.35\textwidth}
        \centering
        \includegraphics[width=0.95\textwidth]{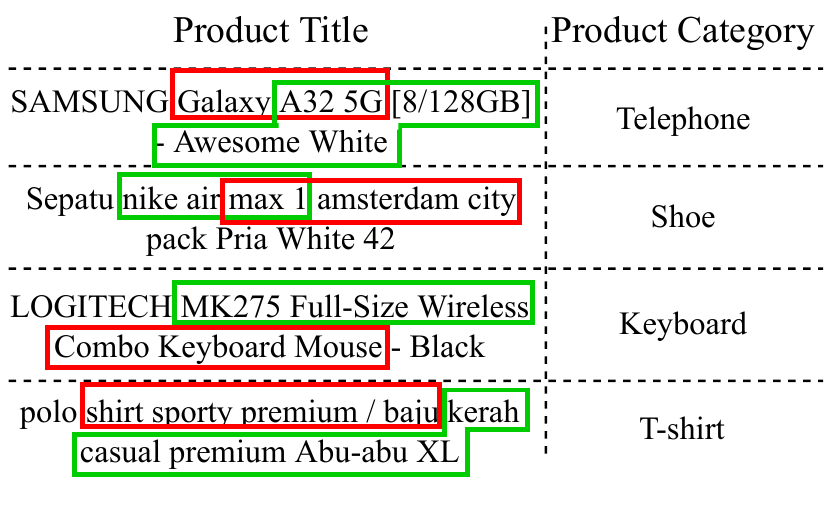}
        \vspace{-3mm}
        \caption{Pretraining data.}
        \label{fig:pretraining_data}
    \end{subfigure}
    \begin{subfigure}{0.25\textwidth}
        \centering
        \includegraphics[width=0.95\textwidth]{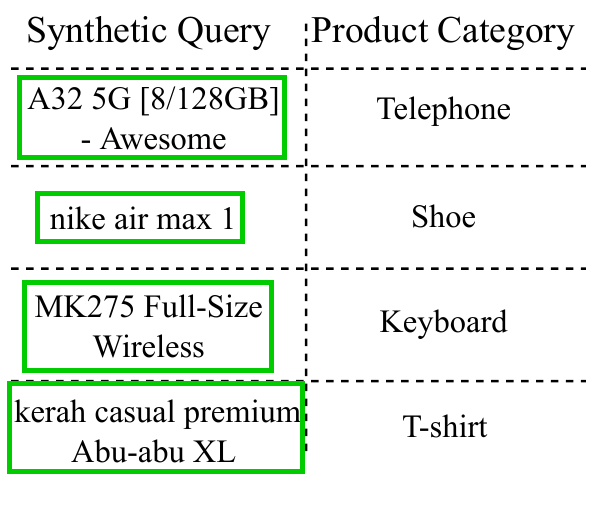}
        \vspace{-3mm}
        \caption{User intent detection.}
        \label{fig:intent_pretrain}
    \end{subfigure}
    \begin{subfigure}{0.36\textwidth}
        \centering
        \includegraphics[width=0.95\textwidth]{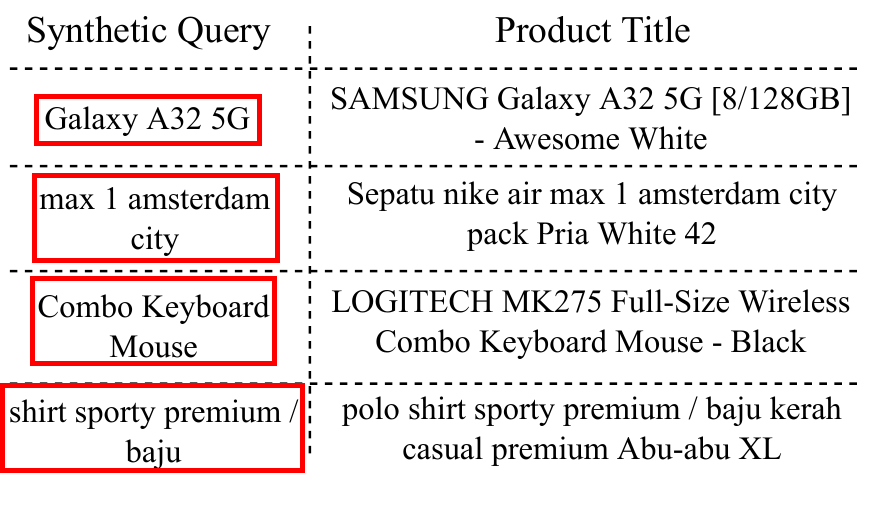}
        \vspace{-3mm}
        \caption{Embedding retrieval.}
        \label{fig:embedding_retrieval}
    \end{subfigure}
    \caption{Illustration of pre-training tasks where we take random substring from item title as synthetic query.}
    \label{fig:pretrain}
\end{figure*}

As shown in Figures~\ref{fig:pretraining_data} and~\ref{fig:intent_pretrain}, the pre-training tasks consist of two sequential ones: 1) \ul{Random Substring Classification (RSC)} refers to the task that we take a random substring from an item title as a synthetic query, and predict the category of the synthetic query as the item's category. Formally, we randomly select a start position of item title according to a uniform distribution between $0$ and title length, and take a substring of random length $l$ sampled again from a uniform distribution from $1$ to a maximum length parameter (5 in our model).  % In addition, we apply a standard BERT encoder to encode fake query to get its [CLS] token embedding and  a softmax classifier to do multi-class classification.
2) \underline{Masked Language Model (MLM)} refers to the standard BERT pre-training task~\cite{devlin2018bert} that randomly masks tokens and learns to recover the masked tokens. We do not adopt the next sentence prediction (NSP) task which is not fit for our scenario, since MLM is more suitable for learning contextualized information. We follow the standard MLM setting which randomly masks 15\% tokens and substitutes 80\% of them with [MASK] token and 10\% with random tokens.

\subsubsection{Fine-tuning.}
The fine-tuning step is very similar to the above pre-training, except for the following two differences: 1) we collect the fine-tuning data by aggregating user click log data, where we collect the most user clicked product categories accounting for up to 90\% of total clicks. Thus, a training instance in the fine-tuning step may consist of a query and several categories, which makes the fine-tuning tasks a mulit-label classification problem instead of a single-label classification problem in the pre-training step.
%data a is a multi-label classification instead of a single-label classification in the pre-training. 
2) We apply the softmax temperature strategy~\cite{hinton2015distilling} to maximize the margin between positive and negative categories. Specifically, we use a temperature 1/3 in our model.
% TODO: rewrite
% In fine-tuning stage, the model is totally warm-started from pre-training model. 
% For a given user's query, the output of fine-tuned model is the query's relevant product categories. Differs from pre-training model, the classification layer is a multi-label formulation, which means there are several possible relevant categories of a query. Moreover, to maximize the margin between positive and negative categories, we apply the softmax temperature strategy~\cite{hinton2015distilling}, which makes a multiplication of logits with a static scalar larger than 1.

% \subsubsection{Details} We use BERT tokenizer for Chinese~\cite{} with vocabulary size as xxx. 

\subsection{Embedding Retrieval}
\label{sec:method:retrieval}
\subsubsection{Problem Formulation.}
Industrial practitioners usually use the two-tower model structure~\cite{zhang2020towards,li2021embedding,liu2021que2search} and approximate nearest neighbor search~\cite{johnson2019billion,guo2020accelerating} libraries to enable fast online retrieval. A typical two-tower model can be formulated as follows. 
\begin{equation}
f(q, s) = Q(q)^\top S(s) \label{eq:scoring}
\end{equation}
where a given query $q$ is input of a query tower $Q$ to generate a query embedding $Q(q) \in \R^{d}$, and an item is input of an item tower $S$ to generate an item embedding $S(s) \in \R^{d}$. Typically, a triplet loss~\cite{hoffer2015deep} including a query, a positive item and a negative item, is optimized during the training.
In our work, pre-training and fine-tuning optimize the same loss function but using different data as described below.

\subsubsection{Pre-training Tasks.}
As shown in Figures~\ref{fig:pretraining_data} and~\ref{fig:embedding_retrieval}, here we use two pre-training tasks similarly as above user intent detection: 1) \underline{Random Substring Retrieval (RSR)} task takes a random substring from an item title as a synthetic query to retrieve the item. An option here is to mask the substring in the item title in order to guide the model to learn semantics instead of word matching. However, in practice we find it makes no noticeable difference in retrieval performance. 2) \underline{MLM} task again refers to the standard BERT pre-training task. As same as user intent detection, we do not include NSP task in our scenario as well.

\subsubsection{Fine-tuning.}
We perform fine-tuning using standard click log data with unique pairs of query and clicked item title. The negative item in the triplet loss is collected in an in-batch negative fashion, which is a common practice in previous works~\cite{zhang2020towards,chang2019pre}.

%% file: 5-experiment.tex
\section{Experiment}
\label{sec:experiment}

\begin{table}[t]
    \vspace{-3mm}
    \centering
    \caption{Dataset Statistics.}
    \small
    \resizebox{0.95\columnwidth}{!}{
    \begin{tabular}{c|cccc}
        \hline
        Dataset   & \# Examples &  \# Queries & \# Items &  \# Categories  \\ \hline
        Pre-training   & 566,161   & --- & 566,161 & 112  \\ 
        User Intent Fine-tuning   & 180,008 & 180,008 & --  & 97 \\ 
        Retrieval Fine-tuning   & 667,665 & 200,001 & 83,672 & --  \\ 
        Overall eval  & 10,000   & 10,000 & 69,855 & 94    \\ 
        Long-tail eval  & 10,000   & 10,000 & 12,312 & 88 \\ \hline
    \end{tabular}}
    \label{tab:dataset}
    \vspace{-3mm}
\end{table}

\subsection{Setup}
\label{sec:setup}

\subsubsection{Dataset.}
Table~\ref{tab:dataset} shows the statistics of our training and evaluation data, which are all collected from user click log on single ``Level-1'' category's items within 60 days. 
% Note that this dataset, open sourced for academic use to reproduce our results and seminate new ideas, is only a subset of our full dataset that is used to train our production model for online A/B test as shown in Section~\ref{sec:online}.
The user intent detection model and semantic retrieval model are pre-trained on the same pre-training dataset, which is collected by item title and product category from our e-commerce commodity pool, but fine-tuned on two different datasets. These two datasets are both collected from user click log with different fields, where user intent detection model uses the search query and product category of clicked item and semantic retrieval model uses the search query and clicked item title.

Then, the two proposed methods are both evaluated on two evaluation datasets, each of which contains 10,000 queries. The overall evaluation dataset contains randomly sampled queries and the long-tail evaluation dataset contains only long-tail queries to measure the model performance on them.
% The user intent detection model and semantic retrieval model are pre-trained on the same pre-training dataset, which contains 566,161 items within 112 categories, finetuned respectively on User Intent Fine-tuning dataset, containing 180,008 queries, and Retrieval Fine-tuning dataset, containing 667,665 query item pairs, and evaluated on the same Overall Evaluation and Long-tail evaluation dataset, which two both contains 10,000 queries.
% For evaluation, we again use the same two datasets, with randomly sampled 10,000 queries and 10,000 long-tail only queries, for both models. 
Note that this dataset, open sourced for the sake of reproducibility of our work or other academic studies, is only a subset of the full dataset used to train the online model in Section~\ref{sec:online}.

\subsubsection{Metrics.}
Our models are evaluated by the metrics of  precision (P), recall (R), f1 score (F1), and the normalized discounted cumulative gains (NDCG or N) for user intent detection task, and precision at top k (P@k) and recall at top k (R@k) for semantic retrieval task.
% Intuitively, precision measures the accuracy of predicted product category of user intent or retrieved items, recall measures the proportion of correctly predicted product categories or correctly retrieved items out of true labels, and normalized discounted cumulative gains is a conventional metric for relevance ranking~\cite{}.
Intuitively, precision measures the accuracy of predicted query categories of user intent or retrieved items, recall measures the proportion of correctly predicted categories or retrieved items out of true labels, and normalized discounted cumulative gains is a conventional metric for ranking tasks, as well as extreme multi-label classification problems~\cite{liu2017deep}.

\subsubsection{Baselines.} 
We compare our customized pre-trained model with three baselines:
\begin{itemize} 
    \item \emph{No pre-train} stands for the model trained directly on the fine-tuning dataset without any pre-training stage. Since the original pre-trained BERT model with 12 layers is infeasible for CPU serving, but only feasible for expensive GPU serving, we explore the variant with 4-layer smaller BERT encoder. Here we conduct 4 and 12 layers \emph{No pre-train} model  experiments for a fair comparison.
    % \item \emph{MLM} stands for the model pre-trained only by the MLM stask, and then finetuned on our dataset.
    \item \emph{BERT-zh} stands for the official pre-trained BERT Chinese model~\cite{devlin2018bert} fine-tuned directly on our dataset.
    \item \emph{Full String Classification (FSC)} stands for the model pre-trained with full item title instead of the random substring, which only suits the intent classification model. Here we only conduct the experiment with 12-layer BERT encoder.
\end{itemize}
Note that all models are optimized by AdamW~\cite{loshchilov2018decoupled} optimizer, and trained with weighted decayed learning rate from 1e-4 and batch size of 1024.

% \begin{CJK*}{UTF8}{gbsn}
% % \linespread{1.2}
% \begin{table*}[t]
% \centering
% \caption{Good cases in semantic embedding retrieval.}
% \label{tab:good_cases_retrieval}
% \setlength{\tabcolsep}{4mm}
% \resizebox{\textwidth}{!}{
% \begin{tabular}{cccc}
% \hline
% Query & RSR+MLM top 1 cases & BERT-zh top 1 cases & Related pre-training titles \\ %& Query in Wikipedia? & \#Pre-train Data \\ 
% \hline
% \specialrule{0em}{2pt}{0pt}
% 卡农线3518 &
% ...卡农线2米 QS3518T2... &
% ...卡侬头音频线... &
% ...卡农线1米 QS3518T2...
% \\
% (XLR line 3518) & 
% (...XLR line 2 meters QS3518T2...) & 
% (...XLR head audio cable...) &
% (...XLR line 1 meters QS3518T1...)
% \\
% \hline
% \specialrule{0em}{2pt}{0pt}
% 惠普953xl墨盒原装 & 
% ...953XL墨盒4色装... &
% 惠普 (HP) 72号墨盒3WX08A...  &
% ...惠普953XL墨盒HP...
% \\ (original HP 953xl ink cartridge) &
% (...953XL ink cartridge 4 colors...) &
% (Hewlett-Packard (HP) No. 72 ink cartridge 3WX08A...) &
% (...Hewlett-Packard 953XL ink cartridge HP...)
% \\
% \hline

% \specialrule{0em}{2pt}{0pt}
% 得力涂改纸 &
% ...得力(deli)...强力附着改正带... &
% 得力 油漆笔补漆笔... &
% ...得力（deli）... 改正带 涂改带8145...
% \\
% (deli correction paper) & 
% (...deli... strong attachment correction tape...) & 
% (deli paint pen touch-up pen...) &
% (...deli... correction tape correction tape 8145...)
% \\
% \hline

% \end{tabular}}
% \vspace{-2mm}
% \end{table*}
% \end{CJK*}

\begin{CJK*}{UTF8}{gbsn}
\linespread{1.2}
\begin{table*}[ht!]
\centering
\caption{Good cases in user intent detection and embedding retrieval.}
\resizebox{0.95\textwidth}{!}{
\begin{threeparttable}[b]
\begin{tabular}{c|cccc}
\hline
  & Query & RSX+MLM & BERT-zh & Related Pre-training Title \\ 
\hline
\specialrule{0em}{2pt}{0pt}
\multirow{4}{*}{User Intent Detection} & 秀才装 & 古装 & 中国历史
& ...男式中国风秀才装演出服蓝色...
\\
 & (Xiucai dress) \tnote{1} & (Ancient Costume) & (Chinese History) & 
(...men's blue Chinese style Xiucai dress...)
\\
\cline{2-5}
\specialrule{0em}{2pt}{0pt}
& 房子出售
& 房子 & 礼品, 装修设计
& ...精装修两室新装修房子出售...
\\
& (house for sale) & (House) & (Gifts, Decoration)  & (...house with two bedrooms ... for sale...) 
\\
\hline
\specialrule{0em}{2pt}{0pt}
\multirow{4}{*}{Embedding Retrieval Recall@1 Case} & 卡农线3518 &
...卡农线2米 QS3518T2... &
...卡侬头音频线... &
...卡农线1米 QS3518T2...
\\
 & (XLR line 3518) & 
(...XLR line 2 meters QS3518T2...) & 
(...XLR head audio cable...) &
(...XLR line 1 meters QS3518T1...)
\\
\cline{2-5}
\specialrule{0em}{2pt}{0pt}
& 惠普953xl墨盒原装 & 
...953XL墨盒4色装... &
惠普 (HP) 72号墨盒3WX08A...  &
...惠普953XL墨盒HP...
\\ 
& (original HP 953xl ink cartridge) &
(...953XL ink cartridge 4 colors...) &
(Hewlett-Packard (HP) No. 72 ink cartridge 3WX08A...) &
(...Hewlett-Packard 953XL ink cartridge HP...)
\\
\hline
\end{tabular}
\begin{tablenotes}
       \item [1] Xiucai is an ancient Chinese academic degree.
\end{tablenotes}
\end{threeparttable}}
\label{tab:good_case}
\end{table*}
\end{CJK*}

\subsection{User Intent Detection}
\label{sec:classification}

We compare the performance of our proposed method with all baseline models in Table~\ref{tab:uit_metrics}, where we can observe that the different versions of our proposed methods, by varying pre-training tasks (\emph{RSC}, \emph{RSC+MLM}) and by varying the network depth (4 and 12 layers), all significantly improve the baseline methods, \emph{No pre-train} and \emph{BERT-zh}. 
Specifically, we can make the following observations: 1) our 12-layer's \emph{RSC+MLM} model improves the baseline \emph{no pre-train} model by 10.4\% in F1 and 6.0\% in NDCG and the baseline \emph{BERT-zh} model by 4.6\% in F1 and 3.3\% in NDCG, on the overall dataset. Note that these three models share exactly the same network structure and only differ in training methods. Thus, these experimental results show that the pre-training tasks are extremely necessary for obtaining a state-of-the-art user intent detection model, and more importantly, the carefully designed pre-training task \emph{RSC} is more proper than the official pre-trained BERT in our scenario of user intent detection task in e-commerce search. We believe this is due to the highly different text in e-commerce data from the standard language in Wikipedia where the official BERT Chinese model is trained from.
% 2) The proposed methods achieve even larger improvements on the long-tail dataset. We believe that the \emph{RSC} pre-training task samples a few training examples to learn the semantics of long-tail queries, better than almost no training examples if without pre-training.
2) The proposed methods achieve even larger improvements on the long-tail dataset. We believe that the sampled queries by \emph{RSC} task could potentially mock the long-tail queries with variant text format.
3) The computationally efficient 4-layer model, though slightly worse than 12-layer model, still outperforms \emph{BERT-zh} model and \emph{No pre-train} model by large gains. Thus, it is a practical trade-off to deploy the 4-layer model in our online production system.
4) By comparing the \emph{RSC+MLM}, \emph{MLM} and \emph{RSC} models, we can see that \emph{RSC+MLM} improves \emph{MLM} by 7.7\% in F1 and 4.0\% in NDCG, but \emph{RSC+MLM} improves \emph{RSC} by only 0.5\% in F1 and -0.1\% in NDCG, on the overall dataset. These results indicate that our proposed pre-training task \emph{RSC} is extremely vital for significant improvements and the standard \emph{MLM} used in the original BERT pre-training does not help much for our task.
5) \emph{RSC} improves \emph{FSC} by 3.6\% in F1 and 2.6\% in NDCG on the overall dataset. We believe this result benefits from more closed length distributions between query and item title.

We illustrate a few good cases in Table~\ref{tab:good_case}. As we can see, a query ``Xiucai dress'' is wrongly predicted as the category ``Chinese History'' by BERT-zh model, while correctly categorized into the category ``Ancient Costume'' by our RSC+MLM model, which potentially learns from the title text where ``Xiucai dress'' co-occurs with other clothes related words.  Again, we now can conclude that, due to the different text distribution between Wikipedia and e-commerce data, a customized pre-training task instead of the official pre-trained BERT-zh model is essential for e-commerce user intent detection.

\begin{table}[t]
\centering
\caption{User intent detection comparative results with baseline methods.}
\label{tab:uit_metrics}
\small
\setlength{\tabcolsep}{1.7mm}
\resizebox{\columnwidth}{!}{
\begin{tabular}{c|cccc|cccc}
\hline
\multirow{2}{*}{Pre-train method} & \multicolumn{4}{c|}{Overall} & \multicolumn{4}{c}{Long-tail} \\
\cline{2-9}
 & P & R & F1 & N & P & R & F1 & N \\
\hline
No pre-train (4-layer) & 0.664 & 0.860 & 0.668 & 0.751 & 0.581 & 0.947 & 0.641 & 0.755 \\
No pre-train (12-layer) & 0.680 & 0.860 & 0.683 & 0.757 & 0.595 & 0.939 & 0.656 & 0.760 \\
MLM (12-layer) & 0.716 & 0.865 & 0.710 & 0.777 & 0.640 & 0.950 & 0.696 & 0.787  \\
BERT-zh (12-layer) & 0.765 & 0.855 & 0.741 & 0.784 & 0.670 & 0.939 & 0.718 & 0.786 \\
FSC+MLM (12-layer) & 0.772 & 0.856 & 0.751 & 0.791 & 0.699 & 0.946 & 0.748 & 0.805 \\
RSC (12-layer) & 0.808 & 0.868 & 0.782 & $\mathbf{0.818}$ & 0.721 & $\mathbf{0.956}$ & 0.770 & $\mathbf{0.831}$ \\
RSC+MLM (4-layer) & 0.782 & $\mathbf{0.870}$ & 0.765  & 0.808& 0.704 & 0.954 & 0.754 & 0.816 \\
RSC+MLM (12-layer) & $\mathbf{0.818}$ & 0.864 & $\mathbf{0.787}$ & 0.817 & $\mathbf{0.737}$ & 0.951 & $\mathbf{0.782}$ & 0.829 \\

\hline
\end{tabular}}
\vspace{-2mm}
\end{table}

% small dataset result
\begin{table}[t]
\centering
\caption{Semantic embedding retrieval comparative results with baseline methods.}
\label{tab:retrieval_metrics}
\small
\setlength{\tabcolsep}{1mm}
\resizebox{\columnwidth}{!}{
\begin{tabular}{c|cc|cc|cc|cc}
\hline
\multirow{2}{*}{Pre-train method} & \multicolumn{4}{c|}{Overall} & \multicolumn{4}{c}{Long-tail} \\
\cline{2-9}
 & R@50 & P@50 & R@100 & P@100 & R@5 & P@5 & R@10. & P@10 \\
\hline
No pre-train (4-layer) & 0.6342 & 0.2103 & 0.7489 & 0.1460 & 0.6295 & 0.1831 & 0.7406 & 0.1164 \\
No pre-train (12-layer) & 0.6415 & 0.2145 & 0.7531 &  0.1481 & 0.6342 & 0.1863 & 0.7480 & 0.1188 \\
MLM (12-layer) & 0.6416 & 0.2159 & 0.7590 & 0.1497 & 0.6445 & 0.1890 & 0.7555 & 0.1199 \\
BERT-zh (12-layer) & 0.6575 & 0.2185 &  0.7689 & 0.1508 & 0.6603 & 0.1943 &  0.7649 & 0.1215 \\
RSR (12-layer) & $\mathbf{0.6685}$ & 0.2216 & $\mathbf{0.7792}$ & $\mathbf{0.1526}$ & $\mathbf{0.6730}$ & 0.1982 & 0.7806 & 0.1245 \\
RSR+MLM (4-layer) & 0.6604 & 0.2189 & 0.7703 & 0.1506 & 0.6690 & 0.1964 & 0.7713 & 0.1223 \\
RSR+MLM (12-layer) & 0.6682 & $\mathbf{0.2219}$ & 0.7786 & 0.1525 & 0.6728 & $\mathbf{0.1990}$ & $\mathbf{0.7806}$ & $\mathbf{0.1247}$ \\
\hline
\end{tabular}}
\vspace{-2mm}
\end{table}

\subsection{Embedding Retrieval}
\label{sec:embedding retrieval}
Table~\ref{tab:retrieval_metrics} presents the performance comparisons between baseline models and our customized pre-training models, measured by R@k and P@k with k=50 and 100 for overall dataset, while k=5 and k=10 for long-tail dataset since long-tail queries have less clicked items in our training dataset. 
Similar conclusions can be made as the above user intent detection model: the necessity of customized pre-training tasks, larger improvement on long-tail queries, and practical tradeoff of the 4-layer model. The only difference here we observe is that the \emph{RSR + MLM} task seems not helping much on top of the \emph{RSR} task, even though the standalone version \emph{MLM} slightly improves \emph{no pre-train} model.
This result actually coincides with previous findings~\cite{chang2019pre} that \emph{MLM} pre-training task does not help much for embedding retrieval.

Table~\ref{tab:good_case} shows a few good cases which benefit from \emph{RSR} task. For instance, ``XLR line 3518'' is a long-tail query which consists of product and model words. Intuitively, \emph{RSR} task facilitates the pre-training model with the key attribute information taken from item title, which enables the fine-tuned model to retrieve items with correct attributes.

% \begin{CJK*}{UTF8}{gbsn}
% \linespread{1.2}
% \begin{table*}[th!]
% \centering
% \caption{Good cases in user intent detection and embedding retrieval.}
% \resizebox{0.95\textwidth}{!}{
% \begin{threeparttable}[b]
% \begin{tabular}{c|cccc}
% \hline
%   & Query & RSX+MLM & BERT-zh & Related Pre-training Title \\ 
% \hline
% \specialrule{0em}{2pt}{0pt}
% User Intent Detection & 秀才装 & 古装 & 中国历史
% & ...男式中国风秀才装演出服蓝色...
% \\
%  & (Xiucai dress) \tnote{1} & (Ancient Costume) & (Chinese History) & 
% (...men's blue Chinese style Xiucai dress...)
% \\
% \hline
% \specialrule{0em}{2pt}{0pt}
% Embedding Retrieval & 卡农线3518 &
% ...卡农线2米 QS3518T2... &
% ...卡侬头音频线... &
% ...卡农线1米 QS3518T2...
% \\
%  Recall@1 Case & (XLR line 3518) & 
% (...XLR line 2 meters QS3518T2...) & 
% (...XLR head audio cable...) &
% (...XLR line 1 meters QS3518T1...)
% \\

% \hline
% \end{tabular}
% \begin{tablenotes}
%       \item [1] Xiucai is an ancient Chinese academic degree.
% \end{tablenotes}
% \end{threeparttable}}
% \label{tab:good_case}
% \end{table*}
% \end{CJK*}

\subsection{Online A/B Test}
\label{sec:online}
Motivated by achieving real world impact from the very beginning, we conduct A/B test on a leading e-commerce search system, using 20\% of the entire site traffic during a period of 30 days. 
Due to the confidential business information protection, we only report the relative improvements in Table~\ref{tab:ab},
where the online baseline models are BiGRU~\cite{li2019joint} for user intent detection and DSPR~\cite{zhang2020towards} for embedding retrieval. The gross merchandise value (GMV), the number of unique order items per user (UCVR), and the click-through rate (CTR) are significantly improved.

\begin{table}[t]
  \setlength\tabcolsep{14pt}
  \caption{Online A/B test.}
  \label{tab:ab}
  \centering
  \resizebox{0.85\columnwidth}{!}{
  \begin{tabular}{cccc}
    \cline{1-4}
     & GMV & UCVR & UCTR \\
    \cline{1-4}
    Intent Detection & +0.799\% & +1.032\% & +0.811\% \\
    Embedding Retrieval & +0.623\% & +0.291\% & +0.012\% \\
    \cline{1-4}
  \end{tabular}}
  \vspace{-3mm}
\end{table}

%% file: 6-conclusion.tex
\section{Conclusion}
\label{sec:conclusion}

In this paper, we have proposed carefully designed, customized pre-training tasks for two critical modules, user intent detection and embedding retrieval in a leading e-commerce search system, since the e-commerce text data are very different from general corpus such as Wikipedia where the official BERT is trained from. As a result, our customized pre-trained models significantly improve no pre-trained models and outperform the official pre-trained BERT models, on both offline evaluation and online A/B test. 